\RequirePackage{lineno}
\documentclass[aps,prl,twocolumn,superscriptaddress,amsmath,amssymb,preprintnumbers,tightenlines]{revtex4}
\usepackage{graphicx} 
\usepackage{dcolumn} 
\usepackage{xcolor} 
\usepackage{amsmath,amssymb} 
\usepackage{hyperref} 
\usepackage[utf8]{inputenc} 
\usepackage[english]{babel} 
\usepackage{physics}
\usepackage{lineno}
\usepackage{orcidlink}

\begin{document}


	\title{	\quad\\[0.1cm]\boldmath First evidence of $X(3872)\to\pi^0\chi_{c0}(1P)$ and search for $X(3915)\to\pi^0\chi_{c1}(1P)$}
  \author{M.~Abumusabh\,\orcidlink{0009-0004-1031-5425}} 
  \author{I.~Adachi\,\orcidlink{0000-0003-2287-0173}} 
  \author{A.~Aggarwal\,\orcidlink{0000-0002-5623-3896}} 
  \author{H.~Ahmed\,\orcidlink{0000-0003-3976-7498}} 
  \author{Y.~Ahn\,\orcidlink{0000-0001-6820-0576}} 
  \author{H.~Aihara\,\orcidlink{0000-0002-1907-5964}} 
  \author{M.~Akdag\,\orcidlink{0009-0004-3728-1077}} 
  \author{N.~Akopov\,\orcidlink{0000-0002-4425-2096}} 
  \author{S.~Alghamdi\,\orcidlink{0000-0001-7609-112X}} 
  \author{M.~Alhakami\,\orcidlink{0000-0002-2234-8628}} 
% ...
\collaboration{The Belle and Belle II Collaborations}

	\begin{abstract}

		We search for the pionic transitions $X(3872)\to\pi^0\chi_{cJ}(1P)$ $(J = 0,~1,~2)$ and $X(3915)\to\pi^0\chi_{c1}$ in $B^+\to \pi^0\chi_{cJ}K^+$ decays using the Belle and Belle~II data samples collected at the $\Upsilon(4S)$ resonance, corresponding to integrated luminosities of $711~\mathrm{fb}^{-1}$ and $492~\mathrm{fb}^{-1}$, respectively.
		We report the first evidence for the decay $X(3872)\to\pi^0\chi_{c0}$ with a significance of $3.4\sigma$, including systematic uncertainties.
		We measure the product of branching fractions ${\cal B}(B^+\to X(3872)K^+)\times{\cal B}(X(3872)\to\pi^0\chi_{c0})=(20.0\pm6.8\pm2.3)\times10^{-6}$ and the branching fraction ratio ${\cal B}(X(3872)\to\pi^0\chi_{c0})/{\cal B}(X(3872)\to\pi^+\pi^-J/\psi)=2.3\pm0.8\pm0.4$, where the first and second uncertainties are statistical and systematic, respectively.
		The upper limits at 90\% credibility on the products of branching fractions for the $\pi^0\chi_{c1}$ and $\pi^0\chi_{c2}$ modes are $7.5\times10^{-6}$ and $15.3\times10^{-6}$, respectively.
        The corresponding upper limits on the branching fraction ratios relative to the $\pi^+\pi^-J/\psi$ decay are $0.9$ and $1.8$.
		The measured branching fractions for $X(3872)\to\pi^0\chi_{cJ}$ are consistent with several theoretical predictions based on the hadronic molecular interpretation of the $X(3872)$. 
		No significant signal is seen for the $X(3915)\to\pi^0\chi_{c1}$ decay, and we set the 90\% credibility upper limit of ${\cal B}(B^+\to X(3915)K^+)\times{\cal B}(X(3915)\to\pi^0\chi_{c1})<6.6\times10^{-6}$, while the decays for $J=0$ and 2 are forbidden by parity conservation.

	\end{abstract}

	\maketitle

The $X(3872)$ state, also known as $\chi_{c1}(3872)$~\cite{pdg}, is the first discovered and best-studied charmonium-like exotic hadron~\cite{review:2017bmm,review:2019esw,review:2024rdk}.
In the two decades since its observation in $\pi^+\pi^-J/\psi$ final state by the Belle experiment in $B^{+}\to\pi^+\pi^-J/\psi K^+$ decays~\cite{X3872:Belle:observe}, 
many features of the $X(3872)$ have been measured experimentally, including the spin-parity $J^{PC}=1^{++}$~\cite{X3872:LHCb:JPC}, a mass close to the $D^{*0}\bar{D}^0$ threshold, an extremely narrow width~\cite{pdg}, and an isospin-violating decay to $\rho J/\psi$~\cite{X3872:Belle:measure}.
However, despite the accumulated experimental knowledge, the nature of $X(3872)$ remains unclear.

The rates of pionic transitions of $X(3872)$ to $\chi_{cJ}(1P)$ $(J=0,~1,~2)$ (denoted as $\chi_{cJ}$ from here on) are predicted to provide strong discrimination among different theoretical interpretations.
These decay rates are expected to be very small for a conventional $\chi_{c1}(2P)$ state~\cite{theory:Dubynskiy:2008}.
The $\chi_{c1}(2P)\to\pi^0\chi_{c0}$ mode is further suppressed as its amplitude vanishes due to the special form of the gluonic matrix element and the approximation for the Green's function of the heavy quark pair as being proportional to the unit operator (for details, please see Ref.~\cite{theory:Dubynskiy:2008}).
However, if $X(3872)$ is a tetraquark or a hadronic molecule, these pionic transitions could be sizable~\cite{theory:Dubynskiy:2008,theory:Fleming:2008}.
Hence measuring the decay rate of $X(3872)\to\pi^0\chi_{c0}$ can distinguish between the charmonium and molecular/tetraquark interpretations.

The $X(3872)\to\pi^0\chi_{cJ}$ decays have been searched for by the Belle~\cite{X3872:Belle:pi0chic1} and BESIII~\cite{X3872:BESIII:pi0chicj,X3872:BESIII:pi0chic0} experiments.
Only the transition to the $J=1$ state was observed by BESIII~\cite{X3872:BESIII:pi0chicj}, which measured the ratio ${\cal{B}}(X(3872)\to\pi^0\chi_{c1})/{\cal{B}}(X(3872)\to\pi^+\pi^-J/\psi)=0.88^{+0.33}_{-0.27}\pm0.10$.  The first and second uncertainties are statistical and systematic, respectively, a convention used throughout this paper.
This result disfavors the interpretation of $X(3872)$ as the $2^3P_1$ charmonium state, given the tiny partial decay width $\Gamma(X(3872)\to\pi^0\chi_{c1})\sim0.06$ keV predicted in Ref.~\cite{theory:Dubynskiy:2008}.
Upper limits (ULs) on the relative branching fractions to the $\pi^+\pi^-J/\psi$ mode for $J=0$ and $2$ at the 90\% confidence level were set by BESIII at 3.6 and 1.1, respectively~\cite{X3872:BESIII:pi0chic0,X3872:BESIII:pi0chicj}.

Measuring the ratio of partial widths for decays to the $\pi^0\chi_{cJ}$ states, denoted $\Gamma_0:\Gamma_1:\Gamma_2$, provides an additional way to discriminate among different theoretical models~\cite{theory:Dubynskiy:2008,theory:Fleming:2008,theory:Dong:2009,theory:Zhou:2019,theory:Wu:2021}. This ratio is predicted to be approximately $0:2.7:1$ for a charmonium $2^3P_1$ state and $2.88:0.97:1$ for a tetraquark or molecular state~\cite{theory:Dubynskiy:2008}.
The absence of experimental observations for the $J = 0$ and 2 modes motivates further studies to understand the nature of the $X(3872)$.

The $X(3915)$, also known as $\chi_{c0}(3915)$~\cite{pdg}, was first observed in $B\to J/\psi\omega K$ by Belle~\cite{X3915:Belle:observe} and identified as a $J^{PC}=0^{++}$ state~\cite{X3915:BaBar:JPC,X3915:LHCb:JPC}.
The decays $X(3915) \to \pi^{0}\chi_{cJ}$ are forbidden by parity conservation for $J = 0$ and $2$; the $J = 1$ decay is allowed but violates isospin.
The decay $X(3915)\to\pi^0\chi_{c1}$ can be searched for in the $B^+\to\pi^0\chi_{c1}K^+$ decays used for the $X(3872)$ studies.

In this Letter, we report searches for $X(3872)\to \pi^0\chi_{cJ}$ and $X(3915)\to\pi^0\chi_{c1}$ in $B^+\to\pi^0\chi_{cJ}K^+$ decays.
The signal $B^{+}$ mesons are fully reconstructed and the $X$ states are searched for in the $\pi^0\chi_{cJ}$ invariant mass spectrum.
Only the $\pi^0\chi_{c1}$ mode is used for the $X(3915)$ search, since the $J=0$ and $2$ modes violate parity conservation.  
We reconstruct $\chi_{c0}$ in the hadronic final states $\pi^+\pi^-$, $K^+K^-$, and $K^+K^-K^+K^-$, and $\chi_{c1,c2}$ via radiative decays to $\gamma J/\psi(\to\ell^+\ell^-,\ \ell=e$ or $\mu$). 
Charge-conjugate channels are implicitly included throughout this paper.
This study is based on a combined analysis using data samples of $772\times10^6$~$B\bar{B}$ pairs (711~fb$^{-1}$) collected by Belle and $524\times10^6$~$B\bar{B}$ pairs (492~fb$^{-1}$) collected by Belle~II.

Belle~\cite{Belle:detector} operated at the KEKB~\cite{kekb,kekb1} asymmetric-energy $e^{+}e^{-}$ collider with electron(positron)-beam energies of 8.0(3.5)~GeV, while Belle~II~\cite{Belle2:detector} operates at its successor, SuperKEKB~\cite{skekb}, with energies of 7.0(4.0)~GeV.
The Belle~II detector is an upgraded version of Belle, and includes a silicon vertex detector composed of two inner layers of pixel detectors and four outer layers of silicon strip detectors, a central drift chamber, a time-of-propagation detector, an aerogel ring-imaging Cherenkov detector, an electromagnetic calorimeter, and an outer $K_L^0$-muon (KLM) detector instrumented in the iron flux return yoke.

Simulated events are used to optimize event selection criteria and to develop the analysis procedures, including background studies and signal extraction, before examining the data. 
The {\sc evtgen}~\cite{evtgen} and {\sc pythia}~\cite{pythia1,pythia2} software packages are used to generate $e^+e^-\to\Upsilon(4S)\to B\bar{B}$ events.
To simulate signal events, $B^+\to XK^+$ and subsequent $X$ decays are modeled according to their nominal $J^{PC}$ values~\cite{pdg}, and the other $B$ meson decays inclusively.
Separate $B\bar{B}$ samples in which both $B$ mesons decay inclusively are used for background studies.
The inclusive samples of continuum processes $e^+e^-\to q\bar{q}$ ($q= u,d,s,c$) are generated with the {\sc kkmc}~\cite{kkmc} and {\sc pythia} packages.
The final-state radiation in both $q\bar{q}$ and $B\bar{B}$ events is simulated by the {\sc photos}~\cite{photos} package.
The detector responses are modeled using the {\sc geant3}~\cite{geant3} software package for Belle and {\sc geant4}~\cite{geant4} for Belle~II.
We use the Belle~II analysis software framework (basf2)~\cite{basf2} to reconstruct and analyze both Belle and Belle~II data.
The Belle data are converted to the Belle~II format for basf2 compatibility using the B2BII framework~\cite{b2bii}.

All charged tracks used to reconstruct the signal events are required to have transverse and longitudinal projections of the distance of closest approach to the interaction point less than 2 and 4~cm, respectively.
The particle identification (PID) assignment for charged tracks at Belle is derived from the likelihood for each particle hypothesis $i$, ${\cal L}(i)$, based on information from the drift chamber, Cherenkov detector, and time-of-flight detector; for electron identification, the calorimeter is also used; for muon identification, the KLM is also used~\cite{b1PID:Nakano:2002jw,b1EID:Hanagaki:2001fz,b1MID:Abashian:2002bd}.
The $\pi$ and $K$ candidates are identified using a binary likelihood ratio ${\cal L}(h)/[{\cal L}(h)+{\cal L}(h^{\prime})]$, where $h^{(\prime)}$ stands for a $\pi$ or $K$.
The electron identification uses a global likelihood variable defined as ${\cal L}(e)/[{\cal L}(e)+{\cal L}({\rm non-}e)]$.
Muons are identified using ${\cal L}(\mu)/[{\cal L}(\mu)+{\cal L}(\pi)+{\cal L}(K)]$.
In Belle~II, hadrons and leptons are identified using neural network trained probabilities, utilizing information from all relevant subdetectors~\cite{b2hid:Belle-II:2025tpe}.
The PID selection achieves an efficiency of $85\%-97\%$ for Belle and $89\%-99\%$ for Belle~II, varying with the particle species and kinematics.
The kaon misidentification probability for pion tracks is 9\%~(7\%) for Belle (Belle~II), respectively, while the pion misidentification probability for kaon tracks is 8\% for both experiments.
We use photons with a minimum energy of 50~MeV and within a 50~mrad angle of electron tracks to recover the lost energy due to bremsstrahlung for electrons.

Energy clusters in the calorimeter not matching any tracks are used to form photon candidates.
To suppress the misreconstruction of fake photons and beam-background in both Belle and Belle~II, we use two boosted decision tree (BDT) classifiers as detailed in Ref.~\cite{photon:mva} for photon candidates.
Pairs of photons whose energy exceeds 70~MeV are used to reconstruct $\pi^0$ candidates.  They must have an invariant mass within the range (0.12--0.15)~GeV/$c^2$, which corresponds to approximately $\pm 3$ times the resolution ($\sigma$) from the known $\pi^0$ mass~\cite{pdg}.

The $\chi_{c0}$ candidates are formed from pairs of charged hadrons $\pi^+\pi^-$ or $K^+K^-$, or from four charged kaons $K^+K^-K^+K^-$.
To determine the correct reconstruction efficiency in the four-body mode, the contribution of the $\phi$ resonance in $\chi_{c0}$ decays is included in simulation~\cite{pdg}.
$J/\psi$ candidates are reconstructed from lepton pairs with invariant mass in the range $2.95\,(3.03) < M_{e^+e^-(\gamma)}\,(M_{\mu^+\mu^-})< 3.13\,(3.13)\,\mathrm{GeV}/c^2$. They are combined with a photon of energy greater than $250\,\mathrm{MeV}$ to reconstruct $\chi_{c1}$ and $\chi_{c2}$ candidates.
All $\chi_{cJ}$ candidates within the loose mass window (3.2, 3.7)~GeV/$c^2$ are kept for further reconstruction.

The $B^{+}$ candidates are reconstructed by combining the four-momenta of the $\pi^{0}$, $\chi_{cJ}$, and $K^{+}$ candidates. 
We select $B^{+}$ candidates by requiring the beam-energy-constrained mass $M_{\rm bc} \equiv \sqrt{E_{\rm beam}^{2}/c^{4}-p_{B}^{*2}/c^{2}} > 5.273~{\rm GeV}/c^{2}$ and the energy difference $\Delta E \equiv E_{B}^{*}-E_{\rm beam} \in (-40,\,30)~{\rm MeV}$, where $E_{\rm beam}$ denotes the beam energy and $p_{B}^{*}$ and $E_{B}^{*}$ denote the momentum and energy of the reconstructed $B$ candidate in the $e^{+}e^{-}$ center-of-mass (c.m.)\ frame, respectively.
The $B^+$ selection requirements have signal efficiencies of about 97\% for $M_{\rm bc}$ and 90\% for $\Delta E$.
A global decay-chain vertex fit is applied to each selected $B^{+}$ candidate using the TreeFit algorithm~\cite{b2fit}, where the masses of the $B^{+}$, $\pi^{0}$, and $J/\psi$ are constrained to their nominal values to improve the mass resolutions of the $\chi_{cJ}$ and $X$ states~\cite{pdg}.
The minimum $\chi^2$ per degree of freedom of the fit is required to be less than 5.

The helicity angle between the momentum of the $\chi_{cJ}$ and the direction opposite to the momentum of the $B^+$, in the rest frame of $X$, is used to further suppress the combinatorial backgrounds.
The $\pi^0$ momentum in the c.m.\ frame must exceed 0.35(0.25)~GeV$/c$ for the $\chi_{c0}$($\chi_{c1,c2}$) mode to suppress background from low momentum neutral particles and multi-body decays.
The ratio of the second to the zeroth Fox-Wolfram moment for the entire event~\cite{r2} is required to be less than 0.25 to suppress continuum backgrounds for hadronic reconstruction modes of $\chi_{c0}$.
The background components in remaining events are investigated using inclusive simulations via the TopoAna algorithm~\cite{topo}.
To suppress backgrounds from specific intermediate states in the $B^{+} \to \pi^{0}\chi_{c0}K^{+}$ mode, we combine the $K^{+}$ candidate with other particles from the signal candidate and veto events whose invariant masses are consistent with the decays $\bar{D}^{0} \to K^{+}\pi^{-}(\pi^{0})$, $K^{*0} \to K^{+}\pi^{-}$, and $\phi \to K^{+}K^{-}$. The corresponding veto regions are $(1.85,\,1.88)$, $(0.84,\,0.96)$, and $(1.01,\,1.03)\,\mathrm{GeV}/c^{2}$, respectively.
These selections reduce the signal efficiency by less than 2\% while rejecting 43\% and 6\% of background candidates for the pionic and kaonic reconstruction modes of $\chi_{c0}$, respectively.
For the $B^+\to\pi^0\chi_{c1,c2}(\to\gamma J/\psi)K^+$ modes, we reject events with $M(\pi^0\gamma)$ in the $\omega$ mass range to suppress the contribution from $X\to\omega J/\psi$ decay.
We reject events when the daughter photon of $\chi_{c1,c2}$ can form a $\pi^0$ candidate with any other photon recorded in the event.
This veto leads to a 33\%~(37\%) reduction in background with signal loss of 12\%~(20\%) in Belle (Belle~II).

All selection requirements for $\pi^0$ reconstruction (including two BDT classifiers and the photon energy), $B^+$ candidate reconstruction, and background suppression of selected $X(3872)$ candidates, except for the mass window requirements on intermediate states, are optimized separately for each $X(3872)$ decay mode by maximizing the Punzi figure-of-merit~\cite{punzi}, $\epsilon / (3/2 + \sqrt{N_{\mathrm{bkg}}})$.
Here $\varepsilon$ is the reconstruction efficiency from signal simulation, ``3'' is the target statistical significance ($3\sigma$), and $N_{\rm bkg}$ is the expected number of background candidates in the signal region estimated from simulated samples.
After the reconstruction and previous selections are applied, an average $B^+$ candidate multiplicity of 1.14~(1.25)
is found in Belle (Belle~II).
For events with multiple $B$ candidates, we retain the one with the smallest TreeFit $\chi^2$, which selects the correct signal candidate with an efficiency of about 70\% according to Monte Carlo simulation.
The reconstructed $\chi_{cJ}$ mass is required to fall in a signal window, defined as [3.39, 3.44], [3.48, 3.53], and [3.535, 3.585]~GeV/$c^2$ for $J= 0,~1$, and 2, respectively.  These windows correspond to approximately $\pm2.5\sigma$, $(-3\sigma, +2\sigma)$, and $(-2\sigma, +3\sigma)$ around the nominal masses for $J= 0,~1$, and 2.
No peaking background is observed in the $M(\pi^{0}\chi_{cJ})$ spectra in studies of the inclusive simulated samples and the events in the $\Delta E$ sidebands, defined as $(-200,-100)$~MeV and $(100,200)$~MeV.

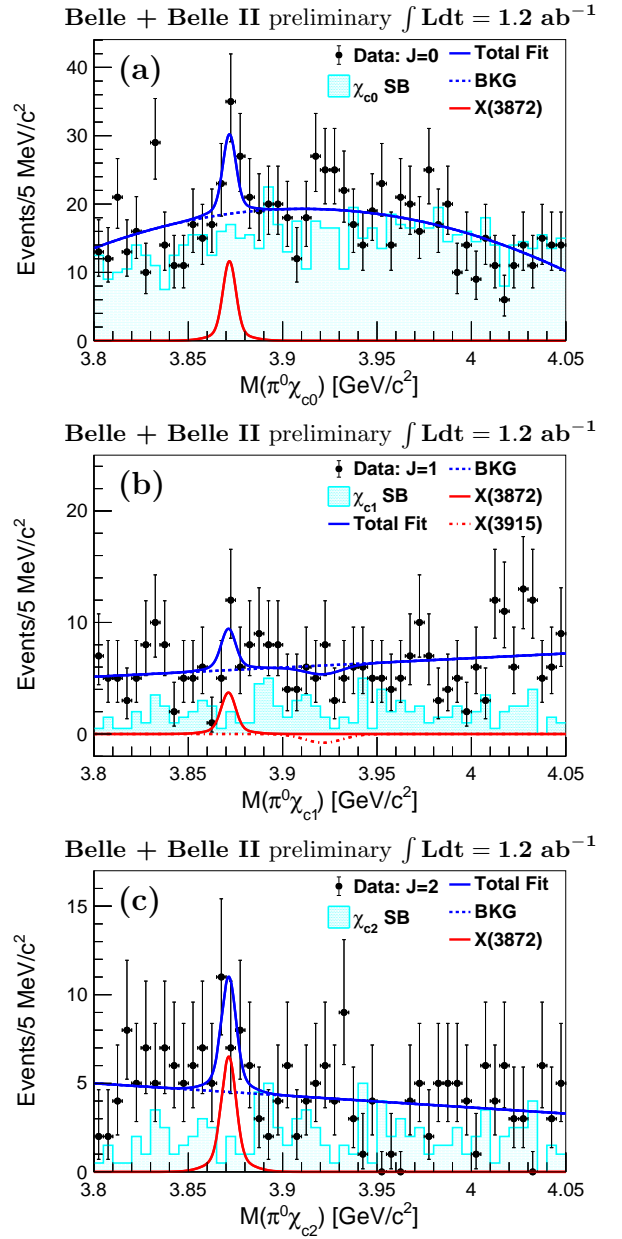
\begin{figure}[!htbp]	
	\centering
	\includegraphics[width=8cm]{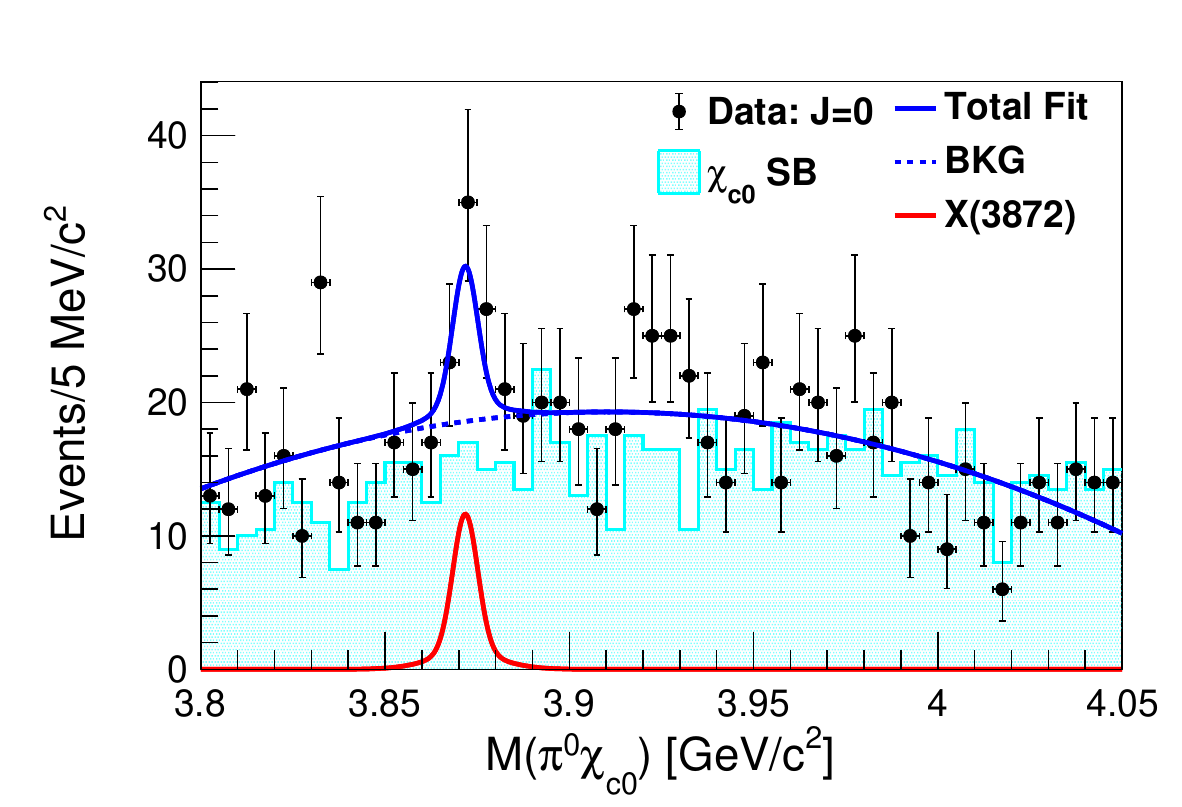} \put(-180,125){\large{\bf (a)}} \put(-200,145){{\bf Belle + Belle II} preliminary $\rm \bf \int Ldt=1.2~ab^{-1}$}
	
	\includegraphics[width=8cm]{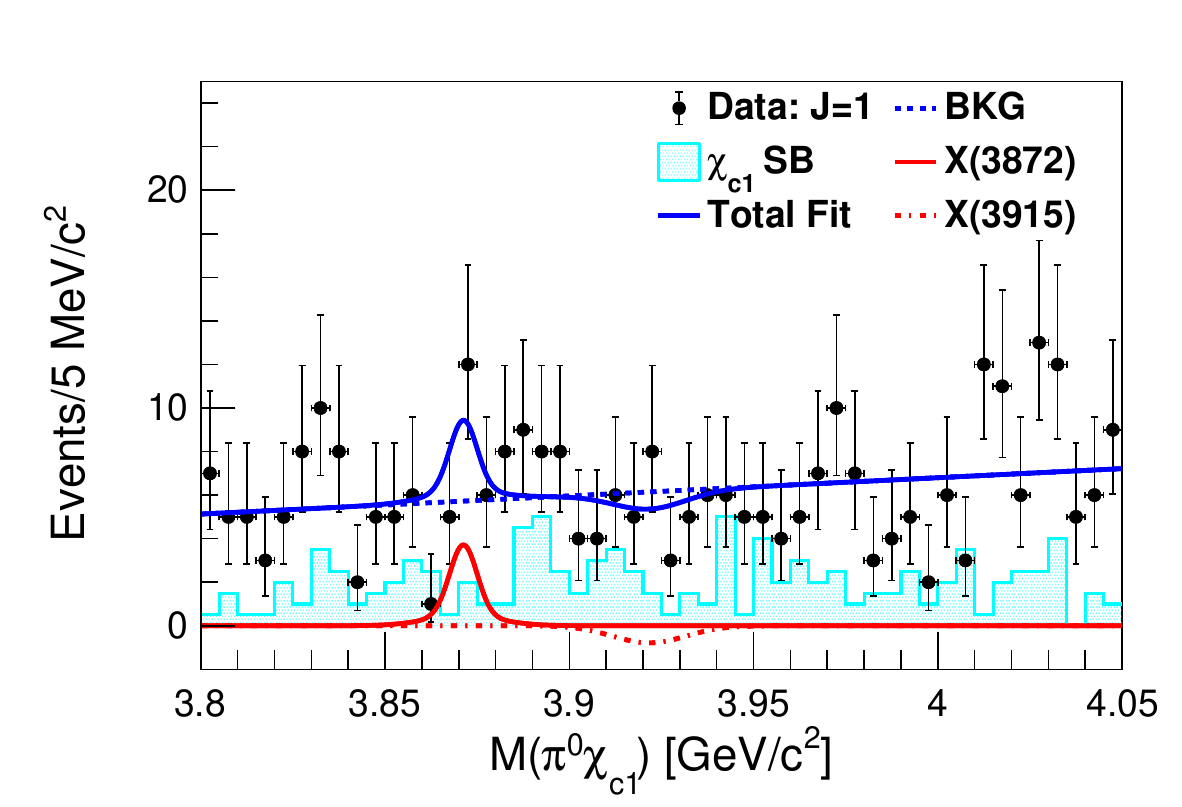} \put(-180,125){\large{\bf (b)}} \put(-200,145){{\bf Belle + Belle II} preliminary $\rm \bf \int Ldt=1.2~ab^{-1}$}
	
	\includegraphics[width=8cm]{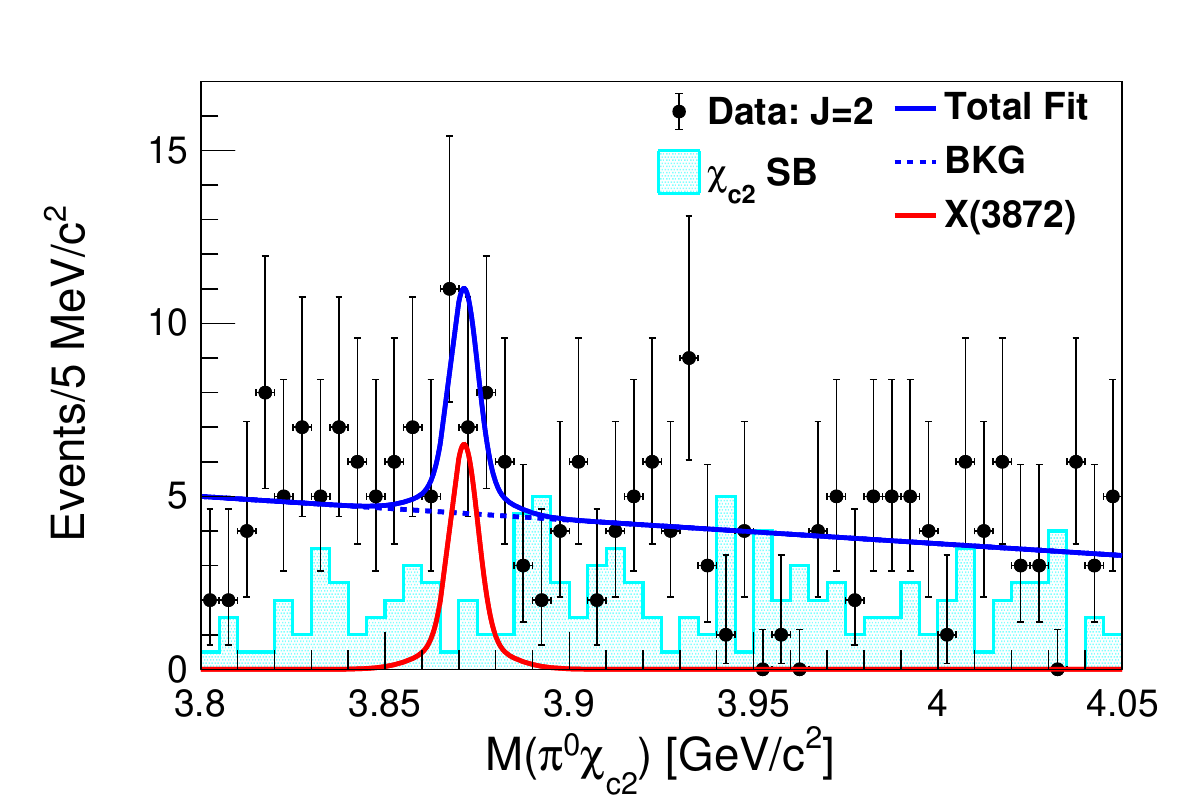} \put(-180,125){\large{\bf (c)}} \put(-200,145){{\bf Belle + Belle II} preliminary $\rm \bf \int Ldt=1.2~ab^{-1}$}

	\caption{The fitted invariant mass spectra of $\pi^0\chi_{cJ}$ for $J$ equal to (a) $0$, (b) $1$, and (c) $2$, respectively, for combined Belle and Belle~II data.
	For each $J$, a simultaneous fit to the $M(\pi^0\chi_{cJ})$ spectra from all reconstruction modes in Belle and Belle~II is performed.
	The dots with error bars represent data samples, blue lines show the total fit results, blue dashed lines show the total fitted backgrounds, and red solid and dashed lines show the fitted $X(3872)$ and $X(3915)$ signals, respectively.
	The cyan histograms show the events in normalized $\chi_{cJ}$ sidebands from data.
	}
	\label{fig:pi0chicj}
\end{figure}

Final distributions of the $\pi^0\chi_{cJ}$ invariant mass for reconstructed $B^+\to\pi^0\chi_{cJ}K^+$ candidates in Belle and Belle~II data are shown in Fig.~\ref{fig:pi0chicj}.
From studies of $\chi_{cJ}$ mass sidebands (SBs) in data and from simulation samples, we find that combinatorial background and three-body $B^+\to\pi^0\chi_{cJ}K^+$ decays both produce a smooth distribution in $M(\pi^0\chi_{cJ})$.
The events from $\chi_{cJ}$ SBs, normalized to the signal window, are shown as cyan histograms in Fig.~\ref{fig:pi0chicj}, where the SB regions are defined as [3.315, 3.365] or [3.465, 3.515]~GeV/$c^2$ for $\chi_{c0}$ and [3.30, 3.35] or [3.60, 3.65]~GeV/$c^2$ for $\chi_{c1,c2}$.
A simultaneous one-dimensional unbinned maximum likelihood fit to Belle and Belle~II data is performed to extract the signal yields of $X(3872)$ and $X(3915)$.
For each $\chi_{cJ}$, all mass spectra from different reconstruction modes in Belle and Belle~II are included in the fit, and the relative contribution to the total signal yield from each mode is fixed to an expectation calculated using efficiencies, intermediate branching fractions, and the numbers of $B^+$ mesons.
For backgrounds, the probability density function (PDF) for each individual reconstructed mode is constructed from a second-order polynomial in the $\pi^0\chi_{c0}$ channel and a straight line in the $\pi^0\chi_{c1,c2}$ channels.
The signal shapes for $X(3872)\to\pi^0\chi_{cJ}$ and $X(3915)\to\pi^0\chi_{c1}$ are derived from signal simulations, where a double-Gaussian function with a common mean is used to model each peak.
All parameters of the signal PDFs are fixed to the values obtained from simulation, while the signal yield for each $X\to\pi^0\chi_{cJ}$ decay mode and all background parameters are allowed to float in the fit.
The simultaneous fit results for $J=0,~1$, and 2 are shown in Figs.~\ref{fig:pi0chicj}(a)--\ref{fig:pi0chicj}(c), respectively.
Validation of the fit using simulated samples confirms that the fit results are unbiased.

\begin{table*}[htbp]
	\caption{
		Results for the search for $X(3872)\to\pi^0\chi_{cJ}$ and $X(3915)\to\pi^0\chi_{c1}$ decays.
		The efficiencies are weighted according to Eq.~(\ref{eq:br}). 
		$\mathcal{B}_X\times\mathcal{B}_J$ stands for the product of branching fractions $\mathcal{B}(B^+\to XK^+)\times\mathcal{B}(X\to\pi^0\chi_{cJ})$.
		The first and second uncertainties are statistical and systematic.
		The ULs are set at 90\% credibility, where the systematic uncertainties are included.
	}
	\centering
	\label{tab:result}
	\footnotesize
	\begin{tabular}{lcccc}
		\hline\hline
		State & \multicolumn{3}{c}{$X(3872)\to\pi^0\chi_{cJ}$} & $X(3915)\to\pi^0\chi_{c1}$ \\
		$J$ & 0 & 1 & 2 &  1 \\
		\hline
		Signal yield & $23.7\pm8.1$ &$8.3\pm5.5$ & $14.8\pm6.0$ &  $-3.9\pm6.6$  \\
		Efficiency(\%) & $6.14\pm0.03$ & $5.23\pm0.03$ & $5.64\pm0.03$ &  $4.94\pm0.03$  \\ 
		Stat. significance (including syst.), in $\sigma$ & 3.6(3.4) & 1.7(1.5) & 3.0(2.6) & - \\
		$\mathcal{B}_X\times\mathcal{B}_J$ ($\times 10^{-6}$) 
		& $20.0\pm6.8\pm2.3$ & $2.9\pm1.9\pm1.0$ & $8.5\pm3.5\pm1.9$  & $-1.5\pm2.5\pm2.7$ \\
		UL: $\mathcal{B}_X\times\mathcal{B}_J$ ($\times 10^{-6}$) 
		& - & 7.5 & 15.3 &  6.6 \\
		\hline\hline
	\end{tabular}
\end{table*}

Based on a signal yield of $23.7\pm8.1$ events, we report the first evidence for the decay $X(3872)\to\pi^0\chi_{c0}$ with statistical significance of 3.6$\sigma$.
The statistical significance is calculated as $\sqrt{-2{\rm ln}({\mathcal{L}}_0/{\mathcal{L}}_{\rm max})}$, where ${\mathcal{L}}_0$ and ${\mathcal{L}}_{\rm max}$ are the maximized likelihoods without and with the signal component.
The total signal yield for $X(3872)\to\pi^0\chi_{cJ}$ ($J=0,~1$ or $2$) is $46.7\pm11.5$ events, corresponding to a statistical significance of $4.3\sigma$.
The significance is evaluated by comparing the likelihoods of simultaneous fits to all $J$ modes with and without the signal components, taking into account the difference in the number of degrees of freedom, which is three~\cite{Wilks:1938dza}.
Including systematic uncertainties, the signal significances of $X(3872)\to \pi^{0}\chi_{c0}$ and $X(3872)\to \pi^{0}\chi_{cJ}$ are $3.4\sigma$ and $3.9\sigma$, respectively.
The significances are taken as the lowest values obtained among all fit variations, as discussed later.
No significant signals are observed for the other $X(3872)$ decay modes or for the $X(3915)\to \pi^{0}\chi_{c1}$ decay, with significances including systematic uncertainties all below $3\sigma$.
The signal yields, efficiencies, and significances are summarized in Table~\ref{tab:result}.

The product of branching fractions $\mathcal{B}(B^+\to XK^+)\times\mathcal{B}(X\to\pi^0\chi_{cJ})$ for each channel is calculated and summarized in Table~\ref{tab:result}, using the formula 
\begin{eqnarray}
\label{eq:br}
&\mathcal{B}(B^+\to XK^+)\times\mathcal{B}(X\to\pi^0\chi_{cJ}) \nonumber\\
&=\frac{N^{\rm fit}}{ \sum_{i}^{\rm Belle\,(II)}\sum_{j}^{N_{\rm mode}} N_{B^+,i}\times\mathcal{B}_{{\rm inter},j}\times\varepsilon_{ij}}.\end{eqnarray}
Here $N_{B^+}$, $\mathcal{B}_{\rm inter}$, and $\varepsilon$ denote the number of $B^+$ mesons, the product of branching fractions of intermediate states, and efficiency, respectively, while $i$ and $j$ are indices corresponding to the experiment (Belle or Belle~II) and reconstruction modes, respectively.
For the decays with significances below $3\sigma$, we also calculate upper limits at 90\% credibility using $\int _0^{x_{\rm UL}}{\cal L}(x)dx\big{/}\int _0^{+\infty}{\cal L}(x)dx = 0.90$, where $x$ is the assumed branching fraction, and ${\cal L}(x)$ is the corresponding maximized likelihood of the ﬁt.
The systematic uncertainties are included in the UL calculation by incorporating the additive uncertainties from the fit procedure and the multiplicative uncertainties from all other sources separately, described below.

Using the previous Belle measurement $\mathcal{B}(B^+\to X(3872)K^+)\times\mathcal{B}(X(3872)\to\pi^+\pi^-J/\psi)=(8.63\pm0.82\pm0.52)\times10^{-6}$~\cite{X3872:Belle:measure}, we calculate the relative branching fractions for each $\pi^0\chi_{cJ}$ final state to the $\pi^+\pi^-J/\psi$ mode.  These branching fraction ratios, denoted as $\Gamma_{J}/\Gamma_{\pi^+\pi^-J/\psi}$, are found to be $2.3\pm0.8\pm0.4$, $0.3\pm0.2\pm0.1$, and $1.0\pm0.4\pm0.2$ for $J=0,~1$, and 2, respectively.
Given that no significant $X(3872)$ signals are observed for the $\pi^0\chi_{c1}$ and $\pi^0\chi_{c2}$ modes, we set ULs at 90\% credibility on $\Gamma_{J}/\Gamma_{\pi^+\pi^-J/\psi}$ as 0.9 and 1.8 for $J=1$ and 2, respectively.
To compare with the theoretical predictions, we calculate additional lower limits (LLs) at 90\% credibility on branching fraction ratios $\Gamma_0/\Gamma_1>2.6$ and $\Gamma_0/\Gamma_2>1.3$. 
Using the Particle Data Group (PDG) average value of $\mathcal{B}(X(3872)\to\pi^+\pi^-J/\psi)=(4.3\pm1.4)\%$~\cite{pdg}, we estimate the absolute branching fraction for $X(3872)\to\pi^0\chi_{c0}$ to be $(9.9\pm4.8)\%$.
The quoted uncertainty includes the statistical and systematic uncertainties of this measurement as well as the uncertainty of $\mathcal{B}(X(3872)\to\pi^+\pi^-J/\psi)$ from the PDG average.
The corresponding ULs on $\mathcal{B}(X(3872)\to\pi^0\chi_{cJ})$ are 3.8\% and 7.6\%
for $J=1$ and $2$, respectively, at 90\% credibility.

We consider systematic uncertainties arising from the fit procedure, detection efficiency (including tracking, PID, and $\pi^0$ and photon reconstruction), best-candidate selection, mass window, vertex fit $\chi^2$ requirement, the size of the Monte Carlo sample,
$N_{B^+}$, and the branching fractions used in this analysis, as described in detail below.
In our measurements, the dominant uncertainties are statistical and those associated with the fit procedure.
The systematic uncertainties are treated differently for the branching-fraction measurements and the ULs on the branching fractions.
For the branching-fraction measurements, the uncertainties from all sources are summed in quadrature.

\begin{table}
	\caption{The absolute systematic uncertainties on the signal yields due to the fit.}
	\centering
	\label{tab:syst:fit}
	\footnotesize
	\setlength{\tabcolsep}{8pt} 
	\begin{tabular}{lcccc}
		\hline\hline
		State & \multicolumn{3}{c}{$X(3872)$} & $X(3915)$ \\
		$J$ & 0 & 1 & 2 	&  1 	 \\
		\hline
		Background polynomial order	& 0.8	& 1.4	& 0.5	& 4.4	 	\\
		Background modeling & 0.1 & 2.1 & 2.6 & 5.7\\
		Fit range 			& 1.4	& 1.0	& 1.4	& 0.5		\\
		Peak position 		& 0.4 	& 0.1	& 0.1	& 0.1	\\
		Mass resolution 	& 0.6 	& 0.2	& 0.8	& 0.1		\\
		\hline
		Total 				& 1.8 	& 2.7 	& 3.1	& 7.2		\\
		\hline\hline
	\end{tabular}
\end{table}

The systematic uncertainty due to the fit procedure is estimated by comparing the nominal fit with alternatives.
We change the order of the background polynomial, add one background component derived from the events in $\Delta E$ sidebands, modify the fit range by $\pm20~{\rm MeV}/c^2$, alter fixed peak positions and mass resolutions considering the data-to-simulation differences from
the control modes of $B^+\to J/\psi(\to\pi^+\pi^-\pi^0)K^+$ and $B^+\to\chi_{c1}[\to\gamma J/\psi(\to\ell^+\ell^-)]K^+$.
For each variation of an individual fit parameter, the change in the signal yield is taken as the corresponding systematic uncertainty, as summarized in Table~\ref{tab:syst:fit}.
The total systematic uncertainty is obtained by adding the individual contributions in quadrature.
The signal significance including systematic uncertainties is defined as the smallest statistical significance obtained among all possible combinations of fit variations.
For the UL determination, we use the fit configuration that yields the largest UL on the signal yield among all combinations of variations. We convolve the corresponding likelihood function with a Gaussian function whose width is set equal to the total systematic uncertainty from all other sources to obtain the final limit.

\begin{table}
	\caption{The relative systematic uncertainties (\%) for branching fraction measurements except those from fitting.}
	\centering
	\label{tab:syst}
	\footnotesize
		\setlength{\tabcolsep}{7.5pt} 
	\begin{tabular}{lcccc}
		\hline\hline
		State & \multicolumn{3}{c}{$X(3872)$} & $X(3915)$ \\
		$J$ & 0 & 1 & 2 	& 1  	 \\
		\hline
		Tracking & 0.8 & 0.8 & 0.8 & 0.8  \\
		Hadron ID & 1.8 & 0.5 & 0.5 & 0.5   \\
		Lepton ID & - & 2.2 & 2.2 & 2.2   \\
		$\pi^0$ reconstruction & 2.8 & 3.3 & 3.3 &  3.2   \\
		Photon reconstruction & - & 1.0 & 1.0 &  1.0  \\
		Best candidate selection & 7.0 & 7.0 & 7.0 & 7.0  \\ 
		Fit $\chi^2$ selection & 1.0 & 1.0 & 1.0 & 1.0\\
		Mass window & 0.5 & 0.5 &0.5 & 0.5 \\
		Simulation sample size & 0.3 & 0.4 & 0.4 &  0.4 \\
		$N_{B^+}$ & 2.2 & 2.2 & 2.2 & 2.2  \\
		$\mathcal{B}_{\rm inter}$ & 3.6 & 3.8 & 4.1 &  3.8   \\
		$\mathcal{B}_X\times\mathcal{B}_{\rm ref}$~\cite{X3872:Belle:measure} & 10.9 & 10.9 &10.9 & -\\
		\hline
		Total for $\mathcal{B}_X\times\mathcal{B}_J$ 	& 9.0 & 9.3 & 9.5 &  9.3 \\
		Total for $\mathcal{B}_J/\mathcal{B}_{\rm ref}$ & 13.8 & 14.0 & 14.1 & - \\
		\hline\hline
	\end{tabular}
\end{table}

All systematic uncertainties except those associated with the fit are summarized in Table~\ref{tab:syst}.
The uncertainty due to tracking efficiency is 0.35\% (0.38\%) per track at Belle (Belle~II), obtained from control samples as described in Ref.~\cite{syst:trk}.
The PID efficiency uncertainties are evaluated as 0.8\% (0.3\%), 0.9\% (0.7\%), 1.6\% (1.2\%), 2.1\% (1.6\%) per track for pions, kaons, electrons, and muons, respectively, in Belle (Belle~II), using control samples of $D^{*+}\to D^0(\to K^-\pi^+)\pi^+$ for hadrons and $J/\psi\to\ell^+\ell^-$ for leptons.
The uncertainties for $\pi^0$ reconstruction are estimated to be 2.3\% (4.8\%) using control samples of $\tau\to\pi^-\pi^0\nu_\tau$ ($D^0\to K^-\pi^+\pi^0$) for Belle (Belle~II), and those for photon reconstruction are 2.0\% (1.1\%) using radiative Bhabha and muon-pair events.
The uncertainty due to the best-candidate selection is estimated as the difference in the mean values of the extracted branching fraction for $X(3872)\to\pi^0\chi_{c0}$ with and without this selection, and the uncertainty due to the vertex fit $\chi^2$ requirement is estimated by changing the selection value by $\pm10\%$.
The uncertainty associated with the mass window requirement is evaluated using the control channel $B^+ \to \chi_{c1}K^+$ by comparing the efficiencies in data and simulation, yielding an uncertainty of 0.5\%.
The uncertainty due to limited simulation sample size is calculated as a binomial uncertainty.
The uncertainties of intermediate state branching fractions are taken from Ref.~\cite{pdg}, and the uncertainty on the reference branching fraction $\mathcal{B}_X\times\mathcal{B}_{\rm ref}\equiv\mathcal{B}(B^+\to X(3872)K^+)\times\mathcal{B}(X(3872)\to\pi^+\pi^-J/\psi)$ is from Ref.~\cite{X3872:Belle:measure}.
The uncertainty on the total number of $B$ meson pairs is 1.4\% (1.5\%) at Belle (Belle~II), and the uncertainty on the ratio of $\Upsilon(4S)$ decaying into a charged $B$ meson pair, $f_{+-}$, is taken from Ref.~\cite{syst:fpm}.
For each source, the uncertainties are summed following the error propagation formula of Eq.~(\ref{eq:br}), assuming that the uncertainties on the reconstruction efficiency in different reconstruction modes are fully correlated and those in different data samples are uncorrelated.
These systematic uncertainties are summed in quadrature. 
For the ratios of $\Gamma_{0}/\Gamma_J$ and $\Gamma_J/\Gamma_{\pi^+\pi^-J/\psi}$, some uncertainties partially cancel, including those from tracking, PID, $\pi^0$, selection criteria, $N_{B^+}$, and $\mathcal{B}(J/\psi\to\ell^+\ell^-)$.

\begin{figure}[htbp]	
	\centering
	\includegraphics[width=9cm]{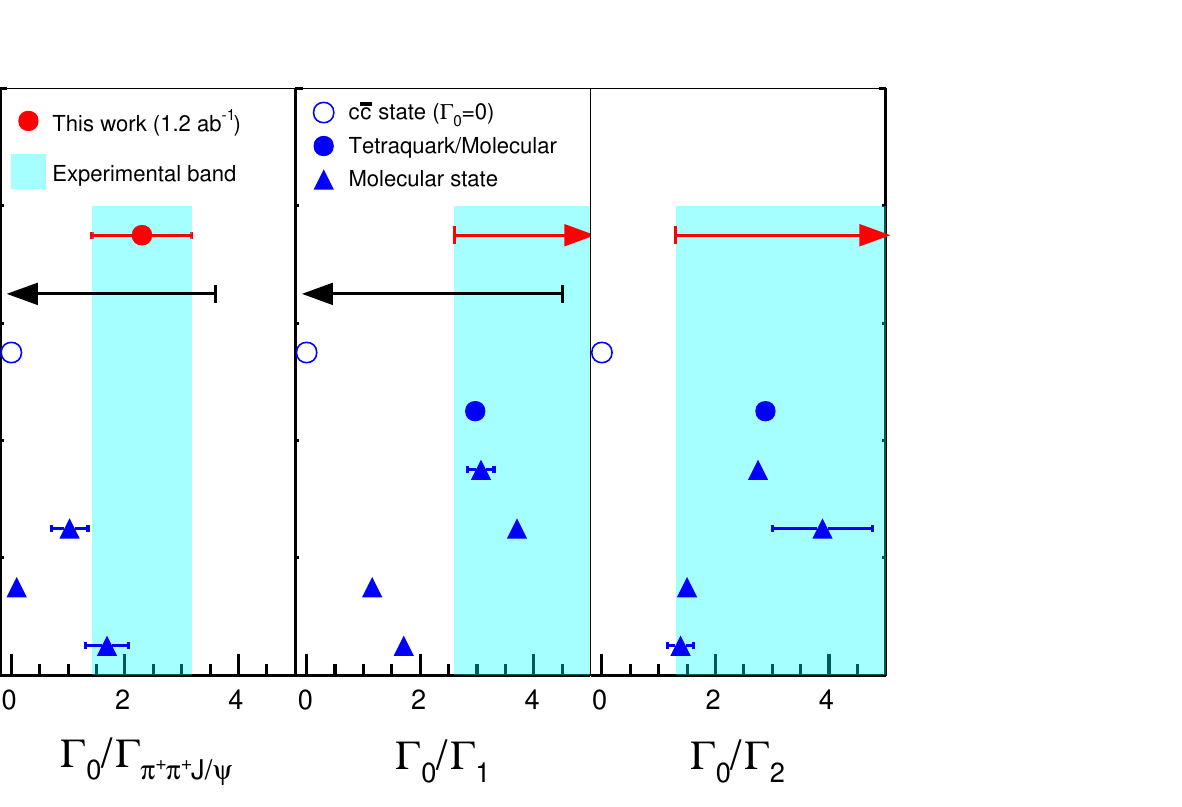}  
	\put(-250,160){{\bf Belle + Belle II} preliminary $\rm \bf \int Ldt=1.2~ab^{-1}$}
	\put(-204,150){\scriptsize{\bf (a)}}		
	\put(-142,150){\scriptsize{\bf (b)}}	
	\put( -80,150){\scriptsize{\bf (c)}}
	\put(-60,121){{\footnotesize This work}}
	\put(-60,109){{\footnotesize BESIII~\cite{X3872:BESIII:pi0chic0}}} 
	\put(-60,97) {{\footnotesize Dubynskiy~\cite{theory:Dubynskiy:2008}}}
	\put(-60,84) {{\footnotesize Dubynskiy~\cite{theory:Dubynskiy:2008}}}
	\put(-60,70) {{\footnotesize Fleming~\cite{theory:Fleming:2008}}}
	\put(-60,58) {{\footnotesize Dong~\cite{theory:Dong:2009}}}
	\put(-60,46) {{\footnotesize Zhou~\cite{theory:Zhou:2019}}}
	\put(-60,34) {{\footnotesize Wu~\cite{theory:Wu:2021}}}
		
	\caption{The comparison of $X(3872)$ partial width ratios (a) $\Gamma_0/\Gamma_{\pi^+\pi^-J/\psi}$, (b) $\Gamma_0/\Gamma_1$, and (c) $\Gamma_0/\Gamma_2$ from measurements in this work, previous BESIII ULs of  $\Gamma_0/\Gamma_{\pi^+\pi^-J/\psi}$ and $\Gamma_0/\Gamma_1$~\cite{X3872:BESIII:pi0chic0}, and theoretical predictions.
	The predictions are derived from a conventional $c\bar{c}$ model in Dubynskiy~\cite{theory:Dubynskiy:2008}, shown as hollow circles, where the $\Gamma(X(3872)\to\pi^0\chi_{c0})$ turns out to be zero, 
	tetraquark/molecular interpretation in Dubynskiy~\cite{theory:Dubynskiy:2008}, shown as filled circles,
	and different approaches in the molecular hypotheses in Fleming~\cite{theory:Fleming:2008}, Dong~\cite{theory:Dong:2009}, Zhou~\cite{theory:Zhou:2019}, and  Wu~\cite{theory:Wu:2021}, shown as triangles. 
	The uncertainty on $\Gamma_0/\Gamma_{\pi^+\pi^-J/\psi}$	combines statistical and systematic contributions.
		The experimental ULs and LLs are shown as arrows.
	}
	\label{fig:result}
\end{figure}
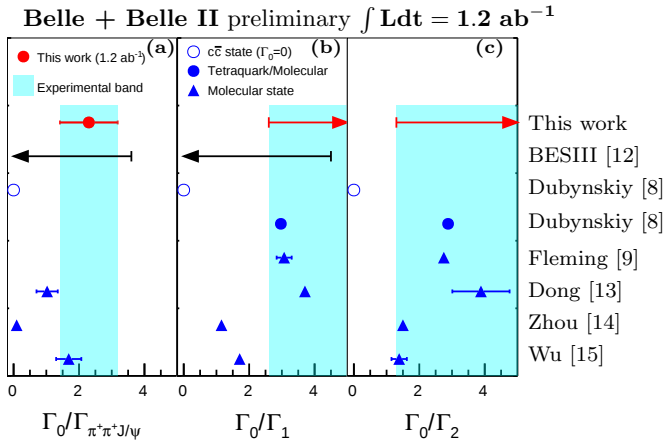

In summary, we report the first evidence for the $X(3872)\to\pi^0\chi_{c0}$ decay with a significance of $3.4\sigma$, including systematic uncertainties, from a study of $B^+\to\pi^0\chi_{cJ}K^+$ decays using Belle and Belle~II data samples corresponding to an integrated luminosity of about $1.2\,\mathrm{ab}^{-1}$. 
The signal significances including systematic uncertainties for $X(3872)\to\pi^0\chi_{c1}$ and $X(3872)\to\pi^0\chi_{c2}$ are $1.5\sigma$ and $2.6\sigma$, respectively.
We measure the product of branching fractions $\mathcal{B}(B^+\to X(3872)K^+)\times\mathcal{B}(X(3872)\to\pi^0\chi_{c0})$ to be $(20.0\pm6.8\pm2.3)\times10^{-6}$.
The corresponding ULs are $7.5\times10^{-6}$ for $\pi^0\chi_{c1}$ and $15.3\times10^{-6}$ for $\pi^0\chi_{c2}$ at 90\% credibility.
The relative branching fractions, also referred to as the partial width ratios, are estimated as 
$\Gamma_{J}/\Gamma_{\pi^+\pi^-J/\psi}=2.3\pm0.8\pm0.4$, $<0.9$, and $<1.8$ for $J=0,~1$, and 2, respectively, and $\Gamma_{0}/\Gamma_{1}>2.6$ and $\Gamma_{0}/\Gamma_{2}>1.3$, where the ULs and LLs are set at 90\% credibility.
These results are statistically consistent with the previous BESIII results~\cite{X3872:BESIII:pi0chicj,X3872:BESIII:pi0chic0}, and update the previous Belle measurements for the $J=1$ channel with tighter ULs~\cite{X3872:Belle:pi0chic1}.
We also estimate the absolute branching fraction of $\mathcal{B}(X(3872)\to\pi^0\chi_{c0})$ to be $(9.9\pm4.8)\%$.
The corresponding ULs of $\mathcal{B}(X(3872)\to\pi^0\chi_{cJ})$ are $3.8\%$ and $7.6\%$ at 90\% credibility for $J=1$ and 2, respectively.

Figure~\ref{fig:result} presents a comparison between theoretical predictions~\cite{theory:Dubynskiy:2008,theory:Fleming:2008,theory:Dong:2009,theory:Zhou:2019,theory:Wu:2021} and our measurements of $\Gamma_{0}/\Gamma_{\pi^+\pi^-J/\psi}$, $\Gamma_{0}/\Gamma_{1}$, and $\Gamma_{0}/\Gamma_{2}$, together with previous BESIII UL results~\cite{X3872:BESIII:pi0chic0}.
If the $X(3872)$ were a charmonium $\chi_{c1}(2P)$ state, the transition amplitude for $X(3872)\to\pi^0\chi_{c0}$ would vanish~\cite{theory:Dubynskiy:2008}.
The observed evidence of the $X(3872)\to \pi^{0}\chi_{c0}$ decay therefore disfavors the pure charmonium interpretation of the $X(3872)$.
Our results are consistent with most predictions based on a molecular scenario.
According to the measured absolute branching fraction, this mode, which exhibits strong isospin violation, may be the largest hidden-charm decay mode of the $X(3872)$ observed so far.

No significant signal for $X(3915)\to\pi^0\chi_{c1}$ is found. 
We set an UL on the product of branching fractions $\mathcal{B}(B^+\to X(3915)K^+)\times\mathcal{B}(X(3915)\to\pi^0\chi_{c1})$ of $6.6\times10^{-6}$ at 90\% credibility.
This UL provides a tighter constraint than the previous Belle result~\cite{X3872:Belle:pi0chic1}.

\begin{acknowledgments}
This work, based on data collected using the Belle II detector, which was built and commissioned prior to March 2019,
was supported by
Higher Education and Science Committee of the Republic of Armenia Grant No.~23LCG-1C011;
Australian Research Council and Research Grants
No.~DP200101792, 
No.~DP210101900, 
No.~DP210102831, 
No.~DE220100462, 
No.~LE210100098, 
and
No.~LE230100085; 
Austrian Federal Ministry of Education, Science and Research,
Austrian Science Fund (FWF) Grants
DOI:~10.55776/P34529,
DOI:~10.55776/J4731,
DOI:~10.55776/J4625,
DOI:~10.55776/M3153,
and
DOI:~10.55776/PAT1836324,
and
Horizon 2020 ERC Starting Grant No.~947006 ``InterLeptons'';
Natural Sciences and Engineering Research Council of Canada, Digital Research Alliance of Canada, and Canada Foundation for Innovation;
National Key R\&D Program of China under Contract No.~2024YFA1610503,
and
No.~2024YFA1610504
National Natural Science Foundation of China and Research Grants
No.~11575017,
No.~11761141009,
No.~11705209,
No.~11975076,
No.~12135005,
No.~12150004,
No.~12161141008,
No.~12405099,
No.~12475093,
and
No.~12175041,
and Shandong Provincial Natural Science Foundation Project~ZR2022JQ02;
the Czech Science Foundation Grant No. 22-18469S,  Regional funds of EU/MEYS: OPJAK
FORTE CZ.02.01.01/00/22\_008/0004632 
and
Charles University Grant Agency project No. 246122;
European Research Council, Seventh Framework PIEF-GA-2013-622527,
Horizon 2020 ERC-Advanced Grants No.~267104 and No.~884719,
Horizon 2020 ERC-Consolidator Grant No.~819127,
Horizon 2020 Marie Sklodowska-Curie Grant Agreement No.~700525 ``NIOBE''
and
No.~101026516,
and
Horizon Europe Marie Sklodowska-Curie Staff Exchange project JENNIFER3 Grant Agreement No.~101183137 (European grants);
L’Institut National de Physique Nucl\'eaire et de Physique des
Particules (IN2P3) du CNRS under Project Identification No.
CNRS-IN2P3-14-PP-033
and L’Agence Nationale de la Recherche (ANR) under Grant No. ANR-23-CE31-
0018 and ANR-25-CE31-1333 (France);
BMFTR, DFG, HGF, MPG, and AvH Foundation (Germany);
Department of Atomic Energy under Project Identification No.~RTI 4002,
Department of Science and Technology,
and
UPES SEED funding programs
No.~UPES/R\&D-SEED-INFRA/17052023/01 and
No.~UPES/R\&D-SOE/20062022/06 (India);
Israel Science Foundation Grant No.~2476/17,
U.S.-Israel Binational Science Foundation Grant No.~2016113, and
Israel Ministry of Science Grant No.~3-16543;
Istituto Nazionale di Fisica Nucleare and the Research Grants BELLE2,
and
the ICSC – Centro Nazionale di Ricerca in High Performance Computing, Big Data and Quantum Computing, funded by European Union – NextGenerationEU;
Japan Society for the Promotion of Science, Grant-in-Aid for Scientific Research Grants
No.~16H03993,
No.~16H06492,
No.~16K05323,
No.~17H01133,
No.~17H05405,
No.~18K03621,
No.~18H03710,
No.~18H05226,
No.~19H00682, 
No.~20H05850,
No.~20H05858,
No.~22H00144,
No.~22K14056,
No.~22K21347,
No.~23H05433,
No.~26220706,
No.~26400255,
and
No.~26H02056,
and
the Ministry of Education, Culture, Sports, Science, and Technology (MEXT) of Japan;  
National Research Foundation (NRF) of Korea Grants
No.~2021R1-F1A-1064008,
No.~2022R1-A2C-1003993,
No.~RS-2018-NR031074,
No.~RS-2021-NR060129,
No.~RS-2024-00354342,
No.~RS-2025-02219521,
No.~RS-2026-25471491,
No.~RS-2026-25480677,
and
No.~RS-2026-25486791,
Radiation Science Research Institute,
Foreign Large-Size Research Facility Application Supporting project,
the Global Science Experimental Data Hub Center, the Korea Institute of Science and
Technology Information (K26L1M2C3)
and
KREONET/GLORIAD;
Universiti Malaya RU grant, Akademi Sains Malaysia, and Ministry of Education Malaysia;
Frontiers of Science Program Contracts
No.~FOINS-296,
No.~CB-221329,
No.~CB-236394,
No.~CB-254409,
and
No.~CB-180023, and SEP-CINVESTAV Research Grant No.~237 (Mexico);
the Polish Ministry of Science and Higher Education and the National Science Center;
the Ministry of Science and Higher Education of the Russian Federation
and
the HSE University Basic Research Program, Moscow;
University of Tabuk Research Grants
No.~S-0256-1438 and No.~S-0280-1439 (Saudi Arabia);
Slovenian Research Agency and Research Grants
No.~J1-50010
and
No.~P1-0135;
Ikerbasque, Basque Foundation for Science,
State Agency for Research of the Spanish Ministry of Science and Innovation through Grant No. PID2022-136510NB-C33, Spain,
Agencia Estatal de Investigacion, Spain
Grant No.~RYC2020-029875-I
and
Generalitat Valenciana, Spain
Grant No.~CIDEGENT/2018/020;
The Knut and Alice Wallenberg Foundation (Sweden), Contracts No.~2021.0174, No.~2021.0299, and No.~2023.0315;
National Science and Technology Council,
and
Ministry of Education (Taiwan);
Thailand Center of Excellence in Physics;
TUBITAK ULAKBIM (Turkey);
National Research Foundation of Ukraine, Project No.~2020.02/0257,
and
Ministry of Education and Science of Ukraine;
the U.S. National Science Foundation and Research Grants
No.~PHY-1913789 
and
No.~PHY-2111604, 
and the U.S. Department of Energy and Research Awards
No.~DE-AC06-76RLO1830, 
No.~DE-SC0007983, 
No.~DE-SC0009824, 
No.~DE-SC0009973, 
No.~DE-SC0010007, 
No.~DE-SC0010073, 
No.~DE-SC0010118, 
No.~DE-SC0010504, 
No.~DE-SC0011784, 
No.~DE-SC0012704, 
No.~DE-SC0019230, 
No.~DE-SC0021616, 
No.~DE-SC0022350, 
No.~DE-SC0023470; 
and
the Vietnam Academy of Science and Technology (VAST) under Grant
No.~DL0000.05/26-27.

These acknowledgements are not to be interpreted as an endorsement of any statement made
by any of our institutes, funding agencies, governments, or their representatives.

We thank the SuperKEKB team for delivering high-luminosity collisions;
the KEK cryogenics group for the efficient operation of the detector solenoid magnet and IBBelle on site;
the KEK Computer Research Center for on-site computing support; the NII for SINET6 network support;
and the raw-data centers hosted by BNL, DESY, GridKa, IN2P3, INFN, 
and the University of Victoria.

\end{acknowledgments}

\bibliographystyle{apsrev4-1}
\bibliography{Ref}

\end{document}